\documentclass[psfig,useAMS,usenatbib]{mn2e}
\usepackage{aas_macros}
\usepackage{graphicx}
\usepackage{epstopdf}
\usepackage{amssymb}

\title{Systematic Bias in Cosmic Shear: Beyond the Fisher Matrix}
\author[A. Amara \& A. Refregier]
{Adam Amara and Alexandre R{\'e}fr{\'e}gier \\
Service d'Astrophysique, CEA Saclay, Gif sur Yvette, 91191, France}

\date{\today}

\pubyear{2001}

\begin{document}

\maketitle

\label{firstpage}

\begin{abstract}
We describe a method for computing the biases that systematic signals introduce in parameter estimation using a simple extension of the Fisher matrix formalism. This allows us to calculate the offset of the best fit parameters relative to the fiducial model, in addition to the usual statistical error ellipse. As an application, we study the impact that residual systematics in tomographic weak lensing measurements.  In particular we explore three different types of shape measurement systematics: (i) additive systematic with no redshift evolution; (ii) additive systematic with redshift evolution; and (iii) multiplicative systematic.  In each case, we consider a wide range of scale dependence and redshift
evolution of the systematics signal. For a future DUNE-like full sky survey, we find that, for cases with mild redshift evolution, the variance of the additive systematic signal should be kept below $10^{-7}$ to ensure biases on cosmological parameters that are sub-dominant to the statistical errors.  For the multiplicative systematics, which depends on the lensing signal, we find the multiplicative calibration $m_0$ needs to be controlled to an accuracy better than $10^{-3}$. We find that
the impact of systematics can be underestimated if their assumes redshift dependence is too simplistic.
We provide simple scaling relations to extend these requirements to any survey geometry and discuss the impact of our results for current and future weak lensing surveys.
\end{abstract}


\section{Introduction}
\label{intro}
Weak gravitational lensing, or `cosmic shear', is undergoing  a phase of rapid expansion \citep[see][for reviews]{2003ARA&A..41..645R,2006astro.ph.12667M,2003astro.ph.10908H} with many future surveys and instruments being planned (e.g. DUNE\footnote{http://www.dune-mission.net}, PanSTARRS\footnote{http://pan-starrs.ifa.hawaii.edu}, DES\footnote{https://www.darkenergysurvey.org}, SNAP\footnote{http://snap.lbl.gov} and LSST\footnote{http://www.lsst.org}).  Central to the planning and designing of these instruments is our ability to predict the uncertainties that such measurements will achieve on the cosmological parameters.  To this end, the Fisher matrix has become a widely used tool in cosmology for calculating their covariance matrix.  However, a limitation of this approach is that it is only able to account for statistical errors, i.e. ones that cause an enlargement of the error bars, and is not well-suited for treatment of systematic errors, which can introduce biases that move the measured central value relative to its true value.   One approach that is commonly taken to overcome this limitation is to treat the systematic errors in the same way as statistical errors and to marginalise over possible values. This introduction of nuisance parameters, in addition to the cosmological parameters, causes the error ellipses to expand.  A more accurate approach, which we use here, is to directly calculate the bias that the systematic signals will introduce.  This bias will tend to offset the central value to the measurements from the true values, as shown in figure \ref{fig:fig0}.  Since the computations needed for this calculation are very similar to those performed in the standard Fisher matrix analysis, extending the current Fisher matrix analysis to include a calculation of bias is relatively straightforward.

We apply this formalism to study the impact of  residual systematics on tomographic cosmic shear surveys. In particular, we consider systematics arising
in the measurement of galaxy shapes after correction of instrumental effects (such
as the Point Spread Function). We consider both additive and multiplicative systematics
and explore a wide range of scale and redshift dependences.
In an earlier work, \cite{2006MNRAS.366..101H} considered the impact of photometric calibration
errors and power law shape systematics using the Fisher matrix formalism.
A similar bias formalism was introduced by \cite{2005APh....23..369H} and applied to theoretical uncertainties in modeling the matter power spectrum with N-body simulations.  Our work expands upon these earlier works, by appying the bias formalism to a broad set of shape systematics and by studying the joint impact of systematic and statistical errors in current and future surveys.

This paper is organised as follows.  In section 2, we describe the formalism that we use to quantify systematic biases.  In section 3, we apply our formalism to cosmic shear surveys by exploring the effect of three types of shape measurement systematics: (i) additive with no redshift evolution; (ii) additive with redshift evolution; and (iii) multiplicative. For each type, we consider several possibilities for their scale dependence: (i) log-linear systematics; (ii) systematics that have the same shape as the lensing signal; and  (iii) systematics that mimic a small change in the cosmological parameters.  In section 4, we study the impact of the systematics in the design of future surveys. Our conclusions are summarised in section 5.
      
\section{General bias Formalism }
\label{general}

In a weak lensing survey, the observed power spectrum
derived from the shapes of background galaxies is given by
\begin{equation}
C_{\ell}^{\rm obs}=C_{\ell}^{\rm lens}+C_{\ell}^{\rm sys}+C_{\ell}^{\rm noise},
\label{eq:cl_tot}
\end{equation}
where each term corresponds to, respectively, the lensing signal ($C_{\ell}^{\rm lens}$), residual systematics ($C_{\ell}^{\rm sys}$), and noise arising from measurement errors and intrinsic shape noise ($C_{\ell}^{\rm noise}$). An estimator of the weak lensing shear power spectrum can thus be defined as
\begin{equation}
\widehat{C_{\ell}^{\rm lens}}=C_{\ell}^{\rm obs}-C_{\ell}^{\rm noise},
\end{equation}
where it is assumed that the residual systematics is unknown and therefore
uncorrected. The errors of this estimator are given by
\begin{equation}
\Delta C_l=\sqrt{\frac{1}{(2\ell+1) f_{\rm sky}}} [ C_{\ell}^{\rm lens}+C_{\ell}^{\rm sys}+C_{\ell}^{\rm noise}],
\end{equation}
where $f_{\rm sky}$ is the fraction of the sky covered by the survey.

The measurement of this power spectrum can then be used to constrain
a set of cosmological parameters $p_{i}$. For this, we form the usual statistic
\begin{equation}
\chi^2(p)= \sum_{\ell} \Delta C_l^{-2} \left[[\widehat{C_{\ell}^{\rm lens}} - C_{\ell}^{\rm lens}(p)]^2 \right].
\end{equation}
An estimator for the parameters $\widehat{p}_i$ is then defined such that $d \chi^2(\hat{p}_{i})/d p_i=0$.

Neglecting the dependence of the errors $\Delta C_{\ell}$ in the parameters, the covariance matrix of the parameters 
\begin{equation}
{\rm cov}[\hat{p}_i,\hat{p}_j]=\langle ( \hat{p}_i - \langle 
\hat{p}_i \rangle ) ( \hat{p}_j - \langle \hat{p}_j \rangle ) \rangle = (F^{-1})_{ij}
\label{eq:cov}
\end{equation}
is then given by the inverse of the Fisher matrix
\begin{equation}
F_{ij}= \sum_{\ell} \Delta C_l^{-2} \frac{d C_{\ell}^{\rm lens}}{d p_i}
\frac{d C_{\ell}^{\rm lens}}{d p_j}.
\end{equation}
It is also easy to show that, for small residual systematics, the bias of the parameter
estimator is given by
\begin{equation}
b[\hat{p}_i] =  \langle  \hat{p}_i \rangle - \langle p_i^{\rm true} \rangle
= (F^{-1})_{ij} B_j,
\end{equation}
where $p_i^{\rm true} $ is the true value of the parameters, the summation 
convention has been assumed and the bias vector $B_j$ is given by
\begin{equation}
B_{j}= \sum_{\ell} \Delta C_l^{-2} C_{\ell}^{\rm sys} \frac{d C_{\ell}^{\rm lens}}{d p_j}.
\label{eq:bias}
\end{equation}
This simple expression is similar to that for the Fisher matrix and is therefore
a convenient way to evaluate the impact of residual systematics  and has also used by \cite{2005APh....23..369H} and \cite{2006MNRAS.366..101H}. Note that this expression, while derived for the measurement of a single weak lensing power spectrum, is general and can be applied
to any estimation of a model using a $\chi^2$ fit in the presence of residual
(unknown) systematics. In particular, it can be easily generalised for the case
where several power spectra are considered, such as in weak lensing tomography,
and if the power spectrum estimators at different multipoles are correlated.

\begin{figure}
\begin{center}
\includegraphics[width=8.5cm,angle=0]{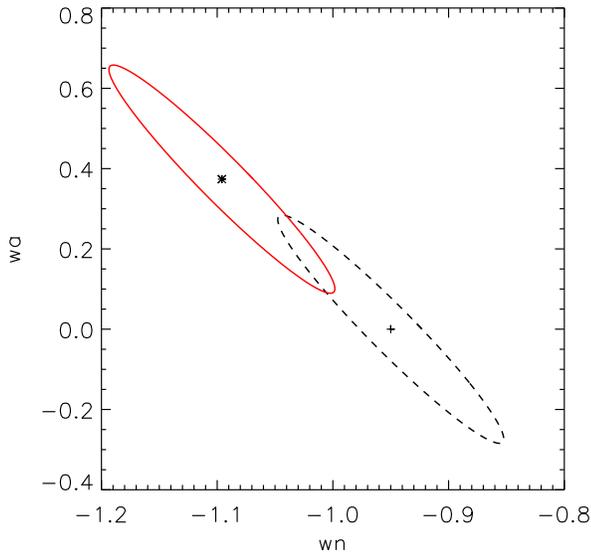}
\caption{Illustration of the distinction between statistical errors which can be estimated using the Fisher matrix and the biasing effect of residual systematics, which can be estimated using the present formalism. The black dashed line shows the results of a Fisher matrix calculation with the cross showing the fiducial model that has been used.  The solid red line shows the error ellipses when the effects of a systematic signal are also included.  We see that the systematic errors can induce a bias that moves the central value relative to the fiducial model.}
\label{fig:fig0}
\end{center}
\end{figure}

Figure \ref{fig:fig0} illustrates the principles of our formalism for the case of the measurement of dark energy equation of state parameters (see detailed discussion below). The black dashed ellipse shows the statistical error estimates for these parameters derived using the fisher matrix. This error ellipse is centred on the fiducial model.  The red error ellipse includes the systematics bias discussed in equation \ref{eq:bias}.  We see that systematics have an additional effect of shifting the centre of the error ellipses away from the fiducial model.

In general, it is convenient to consider the total error covariance matrix
which considers deviations from the true value of the parameters and
is given by
\begin{equation}
{\rm tcov}[\hat{p}_i,\hat{p}_j]=\langle ( \hat{p}_i - p_i^{\rm true} ) ( \hat{p}_j - p_j^{\rm true} ) \rangle
=(F^{-1})_{ij}+b[\hat{p}_i]b[\hat{p}_j].
\end{equation}
This includes both statistical and systematic errors, as opposed to the 
the standard covariance matrix (Eq.~\ref{eq:cov}) which
only considers deviations from the mean values of parameter estimators and
thus only includes statistical errors. In particular, the diagonal elements
of the total error matrix give the Mean Square Error (MSE) of
a parameter $p_i$ 
\begin{equation}
{\rm MSE}[\hat{p}_i]=\sigma^2[\hat{p}_i]+b^{2}[\hat{p}_i],
\label{eq:MSE}
\end{equation}
where the statistical error variance is $\sigma^2[\hat{p}_i]=(F^{-1})_{ii}$. It is this total
error which needs to be minimised when optimising future surveys rather than the statistical error alone. An interesting criterion is to define a tolerance on the systematics such
that they do not dominate over statistical error. This is verified when
\begin{equation}
b[\hat{p}_i] \le \sigma[\hat{p}_i],
\label{eq:sys_tol}
\end{equation}
for all  or a selected subset of the paramters $p_i$. In the following, we will apply this
formalism to cosmic shear and derived the systematics tolerance for future surveys.

\section{Shape Systematics for Weak Lensing Tomography}
\label{shape_syst}

Weak lensing tomography places constraints on cosmological parameters by measuring the statistics of the shear field as a function of redshift. This is done in practice by dividing galaxies into redshift bins and measuring both the auto-correlation of the shear signal within bins as well as the cross-correlation between bins. The shear signal itself is measured using the shapes of distant galaxies. For this purpose, the observed galaxy images must be corrected from the effects of the point spread function (PSF) which is monitored using stars in the image.
Errors in the PSF deconvolution and  galaxy shape estimation induces errors in the measured galaxy shapes, and hence lead to errors in shears. These shape measurement errors generally have spatial correlations (since the PSF itself has a spatial variation) which can be described by its power spectrum $C_{\ell}^{\rm sys}$. In this work, we decompose residual systematic power spectrum into two parts: an additive term (i.e. independent of the lensing signal) and a multiplicative term (i.e. dependent on the lensing signal). Each term and its impact on the systematics power spectrum is described in Appendix\ref{sec:appA}. 

In the following, we consider a  flat cosmological model with 7 parameters listed in table \ref{tbl:stat_err}. In particular, the evolution of the dark energy equation of state parameter is assumed to take the form $w(a)=w_n+(a_n-a)w_a$, with $w_0$, listed
in the table, corresponding to a pivot point of $a_n=1$. For the cosmic shear survey parameters, we focus on the DUNE-like `shallow' survey described in \cite{2006astro.ph.10127A}, namely a 20,000 sq. degree survey containing 35 galaxies per amin$^2$  with $\sigma_\gamma=0.25$ and $z_m=0.9$, with the galaxies divided into 5 redshift bins so that each bin contains the same number of galaxies. We assume that the overall galaxy distribution is given by \cite{1994MNRAS.270..245S},
\begin{equation}
\label{ }
P(z)=z^\alpha \exp \bigg[-\bigg(\frac{z}{z_0}\bigg)^\beta\bigg],
\end{equation}
where we set $\alpha=2$ and $\beta=1.5$.  The median redshift of the survey, $z_m$, is then used to set $z_0 \simeq z_m/1.412$. We use multipoles in the range $10<\ell<2\times10^4$.  The lensing tomography formalism we use is also described in \cite{2004PhRvD..70d3009H} and \cite{2006astro.ph.10127A}. Table \ref{tbl:stat_err} shows the central values of our fiducial model and the marginalised Fisher matrix errors on each of the cosmological parameters for this survey.  These errors come from the same calculation used to produce the results (dashed curve) shown in figure \ref{fig:fig0}.  The errors on the dark energy parameters are sometimes stated in terms of $w_n$ corresponding to the pivot point $a_n$ where $w_n$ and $w_a$ are uncorrelated.  For our survey the marginalised error on $w_n$ is 0.02.

\begin{table*}
  \centering 
\begin{tabular}{c|c|c|c|c|c|c|c}
\hline
Parameter &$\Omega_m$& $w_0$& $w_a$& h& $\sigma_8$& $\Omega_b$& $n$ \\
Central value & 0.28 & -0.95 & 0.00 & 0.72 & 1.0 & 0.046 & 1.0 \\
Marginalised errors   & 0.006 & 0.06 &0.19&17&0.008&4&0.009 \\
\hline
\end{tabular}
  \caption{Cosmology parameters for our fiducial model with marginalised Fisher matrix errors for the survey we consider.}
  \label{tbl:stat_err}
\end{table*}

\subsection{Additive Term}

First, we consider an additive systematic signal, which, by definition is not correlated with the lensing signal.
Such a systematic could, for instance, result from residual errors in the correction of the PSF. This signal
can in general have an arbitrary scale dependence and may also depend on the galaxy redshift.

To quantify the amplitude of the systematic signal it is convenient to consider its variance
\begin{equation}
\label{eq:var}
\sigma^2_{\rm sys} =\frac{1}{2\pi} \int |C_\ell^{\rm sys}|\ell(\ell+1)d\ln\ell,
\end{equation} 
where the absolute value sign is included to account for possible changes of sign.  
To explore the possible scale dependence of the systematic signal, we consider the following three classes
of shape of the systematic power spectrum:
\begin{itemize}
\item{The first is a log-linear systematic:
\begin{equation}
\label{eq:powlaw}
\ell(\ell+1)C_{\ell}^{\rm sys} = A_0\big(n \log_{10} (\ell/\ell_0)+1\big),
\end{equation}
where $\ell_0$ is a reference scale, $n$ a scaling parameter and $A_0$ is a normalisation. This parametrisation allows for the possibility that the residual power spectrum of systematic signal, after correction, be positive or negative and may transition from one to the other.  More specifically, the quantity $\ell(\ell+1)C_{\ell}^{\rm sys} $ scales linearly with $\log \ell$ and goes through the point $\ell(\ell+1)C_{\ell}^{\rm sys}=A_0$ at $\ell=\ell_0$ and has a slope of $nA_0$}.In general, each correlation function could have its own normalisation (i.e. $A_{ij}$, rather than $A_0$). For instance this would be the case if the systematic signal had a redshift dependence.
\\
  \item {Secondly, we explore systematic signals that have the same shape as one of the lensing power spectra, $C_{\ell}^{\rm lens}$.  Since, in this work, the galaxies have been divided into 5 redshift bins, this gives us 15 possible power spectra to investigate:
\begin{equation}
\label{eq:cl}
C_{\ell}^{\rm sys} = A_1C_{\ell}^{ij},
\end{equation}
where $C_{\ell}^{ij}$ is the correlation spectrum between bins i and j, and $A_1$ is a normalisation. Once again each correlation function could have a different normalisation. For instance this would be the case if the systematic signal had a redshift dependence.}\\
 \item {Finally, we explore the systematic shape that should have the greatest impact on our measured cosmological parameter estimation, namely systematics that exactly mimic  the effect of a change in one of the cosmological parameters ($p_\alpha$):
\begin{equation}
\label{eq:dcldp}
C_{\ell}^{\rm sys} = A_2 \frac{dC_{\ell}^{ij}}{dp_\alpha},
\end{equation}
where$A_2$ is another normalisation factor.}
\end{itemize}

\begin{figure}
\begin{center}
\includegraphics[width=8.5cm,angle=0]{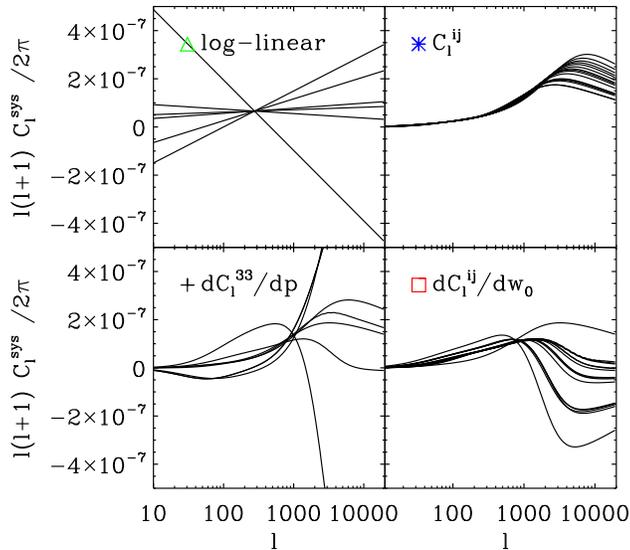}
\caption{Different classes of additive systematic signals we consider. In each case, the systematics power spectra shown have been normalised so that they induce a 
a bias on $w_0$ equal to the systematics error ($b[w_0]=\sigma[w_0]=0.06$).  The top left panel shows examples of the log-linear models for n=-1.4, -1.0, -0.6, -0.2, 0.2 and 0.6; the top right panel shows systematics that have the same shape as the lensing signal (all 15 powerspectra shapes are shown); and the bottom left panel shows signal that have the shape of $dC^{33}_\ell/dp_\alpha$, where $C^{33}_\ell$ is the auto-correlation power spectrum of the third redshift slice and $p_\alpha$ are the parameters in our fiducial model. Finally, the bottom right panel shows systematic signals that have the same functional form as  $dC^{ij}_\ell/dw_0$, where $C^{ij}_\ell$ is the lensing correlation function between bins i and j.}
\label{fig:fig2}
\end{center}
\end{figure}
 
Figure \ref{fig:fig2} shows examples of these different classes of systematic signals.  Each of the systematic signals have been normalised so that they introduce a bias on the equation of state parameter $w_0$ of 0.06, which is the same level as the marginalised statistical error on this parameter calculated from a standard Fisher matrix approach
see Table~\ref{tbl:stat_err}.

\begin{figure}
\begin{center}
\includegraphics[width=8.5cm,angle=0]{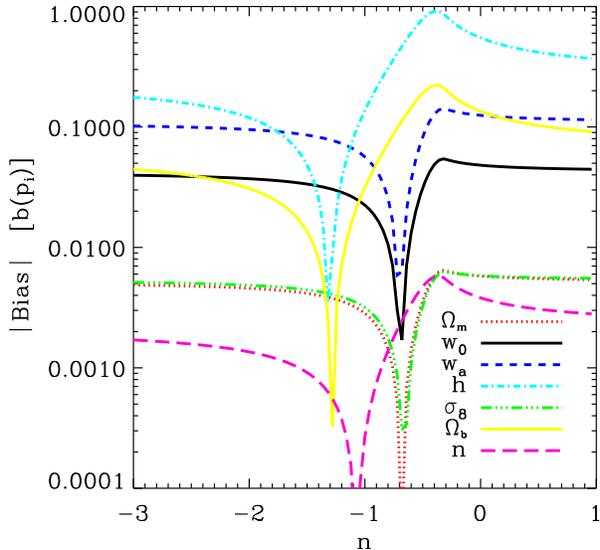}
\caption{The bias of all 7 parameters considered in our fiducial model produced by the log-linear systematic as a function of n, the scaling in equation \ref{eq:powlaw}.  $A_0$ is chosen so that $\sigma_{\rm sys}^2 = 4\times10^{-7}$.}
\label{fig:fig3}
\end{center}
\end{figure}

\subsubsection{Additive term with no redshift evolution}

We first consider additive systematics whose amplitude is independent of redshift.
Figure \ref{fig:fig3} shows the bias introduced to each parameter in our log-linear model (Eq.~\ref{eq:powlaw} and top left panel of figure \ref{fig:fig2})) as function of the scaling parameter $n$. The 7 curves in the figure show the bias for a systematic signal with an amplitude such that $\sigma_{\rm sys}^2=4\times10^{-7}$. As can be seen from equations \ref{eq:bias} and \ref{eq:var}, the bias scales linearly with $\sigma_{\rm sys}^2$. 
The level of the bias depends on the cosmological parameter considered and the absolute value of the bias remains roughly flat with $n$.  The apparent dips around $n=-1$ correspond to a change of sign of the biases (e.g. goes from positive to negative). 
The black-solid and blue-dashed curves show that a log-linear systematic with $\sigma_{\rm sys}^2 = 4\times 10^{-7}$ will introduce a bias of $\sim0.05$ on $w_0$ and $\sim0.1$ on $w_a$, for most values of $n$.   The range of possible bias values is large depending on which cosmology parameter is under scrutiny (note that the $y$-axis of figure \ref{fig:fig3} is on a log scale).  For example, the absolute bias induced in $\Omega_m$  is typically $\sim0.005$ for a wide range of n values.  

\begin{figure}
\begin{center}
\includegraphics[width=8.5cm,angle=0]{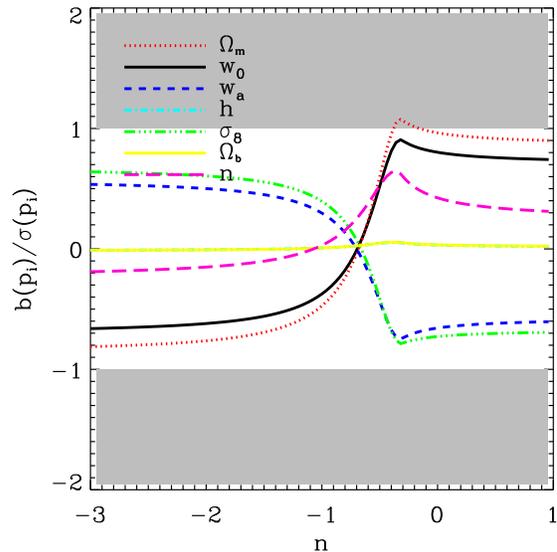}
\caption{The ratio of the systematic bias to the marginalised statistical errors for each of the cosmological parameters as a function of
 $n$, the scaling parameter of the log-linear scaling model. The amplitude has been set to $\sigma_{\rm sys}^2 = 4\times10^{-7}$.  The grey area corresponds to regions where the bias dominates over the statistical errors (i.e. $|b(p_i)/\sigma(p_i)| >1$). We see that it is not always the dark energy equation of state parameters that drive the systematic requirements.}
 \label{fig:fig4}
\end{center}
\end{figure}

Although the value of the bias on the $\Omega_m$ is smaller than that for $w_0$, a more relevant quantity is the size of the bias relative to the statistical errors (which are shown in table \ref{tbl:stat_err}).  Figure \ref{fig:fig4} shows the ratio of the bias to the marginalised Fisher matrix errors of all 7 cosmology parameters for the log-linear model with  $\sigma_{\rm sys}^2=4\times10^{-7}$. Expressed in this manner, we see that many of the cosmological parameters respond to the same extent to the presence of a systematic signal (note that the $y$-axis is now on a linear scale).  We also see more clearly that as we vary $n$, the bias on any given parameter can change sign. 

The grey shaded region in the figure is where the bias dominates over statistical errors, i.e. where $|b(p_i)|/\sigma(p_i) > 1$.  With this in mind, we can now adjust the level of the the systematic ($\sigma_{\rm sys}^2$) and find our tolerance ($\sigma_{\rm sys-tol}^2$), which, from equation \ref{eq:sys_tol}, is the maximum value that $\sigma_{\rm sys}^2$ can have before the potential bias on one of the cosmology parameters breaches our acceptable threshold (i.e. before one of the curves in figure \ref{fig:fig4} enters the grey area).  Figure 4 shows that, for the log-linear systematic, setting a threshold of $\sigma_{\rm sys-tol}^2=4\times10^{-7}$ would be sufficient to meet this criterial for the dark energy parameters.

Appendix \ref{apx:b}, shows the results for the other classes of systematics.  Table \ref{tbl:1} shows the ratio of the bias to marginalised statistical errors for the residual systematic signals shown in the top right panel of figure  \ref{fig:fig2}, table \ref{tbl:2} correspond to the bottom left panel figure  \ref{fig:fig2} and \ref{tbl:3} are the results for the bottom right panel. We see that, if the residual systematic has the same shape as the lensing signal, then a tolerance level of $\sigma_{\rm sys-tol}^2=4\times10^{-7}$ is sufficient to keep $|b(p_i)/\sigma(p_i)| <1$ for the $w_0$ and $w_a$  parameters.  However, table \ref{tbl:3} shows that a systematic level of $\sigma_{\rm sys-tol}^2=4\times10^{-7}$  can lead to a ratio greater than one for $w_0$ for some systematic power spectrum shapes, for instance $C_\ell^sys \propto  dC_\ell^33/dw_0$.  It is also interesting to note the the most stringent constraints on systematics does not always come from dark energy. For instance for a systematic that has a shape $C^{sys}_\ell \propto dC^{45}_\ell/dw_0$, it is the constraint from $\Omega_m$ that would dominate. 

\begin{figure}
\begin{center}
\includegraphics[width=8.5cm,angle=0]{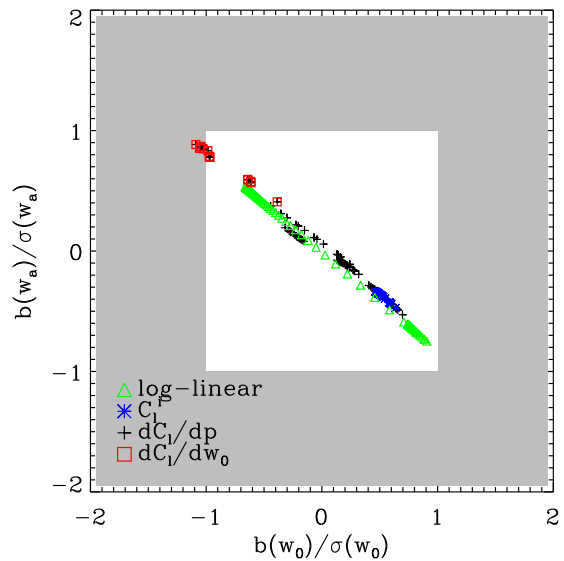}
\caption{Impact of additive systematics (with no redshift evolution) on the dark energy
parameters, $w_0$ and $w_a$, for all classes of scale dependence considered.
As before, the grey area corresponds regions where the bias dominates over the (marginalised) statistical errors.  In all cases, we set $\sigma_{\rm sys}^2 = 4\times10^{-7}$. }
\label{fig:fig5}
\end{center}
\end{figure}

Focusing again on dark energy parameters, figure \ref{fig:fig5} shows the biases (as a fraction of statistical errors) for $w_0$ and $w_a$ for all the additive systematics we have considered.  Once again the grey area shows regions where the bias would dominate over the statistical errors.  We see that a systematic with $\sigma_{\rm sys-tol}^2=4\times10^{-7}$ is not always sufficient (this can also be seen in tables \ref{tbl:2} and \ref{tbl:3}).  Instead a more comfortable tolerance level is $\sigma_{\rm sys-tol}^2 =3\times10^{-7}$, which we can be considered as the target for the control of additive systematics with \it no \rm redshift evolution.

\subsubsection{Additive term with redshift evolution}

The additive term discussed in the previous section was for a systematic signal that is the same for all tomographic power spectra, i.e. independent of the galaxy redshifts.  An extension of the above results is to consider an additive term that evolves with redshift.  This, for example, could be due to the fact that more distant galaxies tend to be smaller than nearby galaxies, making their shape measurements more prone to errors in PSF deconvolution.  We can model such an effect by introducing a simple redshift scaling to the additive part of equation \ref{eq:gamma_sys}:
\begin{equation}
\label{eq:add_sc}
\gamma^{add} = \gamma_0^{add}(1+z_m)^{\beta_a},
\end{equation}
where $z_m$ is the median redshift of a tomographic redshift bin. The systematic power spectrum now depends on the redshift of the two galaxies being correlated and, for simplicity, we assume that this redshift dependence can be separated from the angular dependence such that,
\begin{equation}
C_\ell^{sys}(z_m^i,z_m^j)=(1+z_m^i)^{\beta_a}(1+z_m^j)^{\beta_a}C_\ell^{sys}(z_m^i=0,z_m^j=0),
\end{equation}
where $z_m^i$ and $z_m^j$ are the median redshifts of bins i and j, and $C_\ell^{sys}(z^i=0,z^j=0)$ is the systematic auto-correlation power spectrum of galaxies at redshift z=0. 

\begin{figure}
\begin{center}
\includegraphics[width=8.5cm,angle=0]{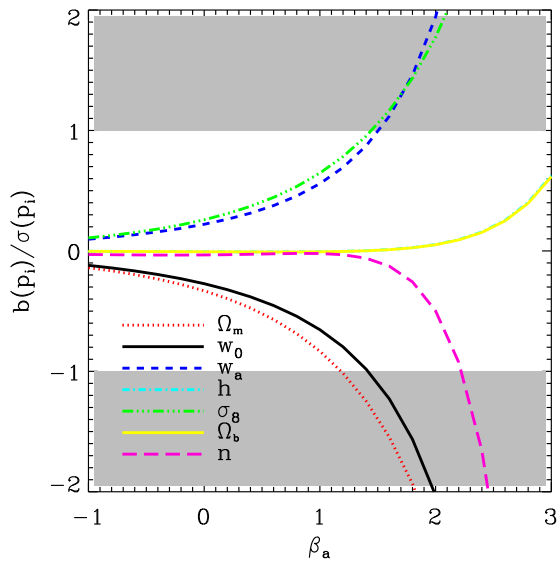}
\caption{The ratio of bias to statistical errors for additive systematics with redshift evolution for the different cosmological parameters. The shape of the systematics has been chosen to be $C^{sys}_\ell \propto dC^{33}_\ell/dw_0$ and $\beta_a$ is the power-law exponent describing the redshift evolution (see text). In all cases, the systematic auto-correlation variance at $z=0$ is set to $\sigma_{\rm sys}^2 = 10^{-7}$. Note that this is a systematic amplitude that is a factor of 4 smaller than that used in earlier figures.}
\label{fig:fig6}
\end{center}
\end{figure}

Figure \ref{fig:fig6} shows the bias (as a fraction of marginalised errors) for different values of $\beta_a$ and for systematic shape of the form $C_\ell^{sys} \propto   dC^{33}_\ell/dw_0$ (one of our worst-case systematic shapes) and an amplitude such that $\sigma^2_{\rm sys}=1\times10^{-7}$.  Once again we see that it is not always the dark energy equation of state parameters that drive the systematic requirements, and that $\Omega_m$ and $\sigma_8$ can be as restrictive as the two $w$ parameters.  We also see large values of $\beta_a$ cause the bias to become very large relative to the statistical errors. For the dark energy parameters, we therefore place a requirement of $\beta_a < 1.5$ and $\sigma^2_{\rm sys-tol}(z^i=0,z^j=0)<1\times10^{-7}$. 

\subsection{Multiplicative Term}

Having investigated the effect of additive systematics, we now investigate the case of multiplicative systematics which are correlated to the lensing signal. This is most easily done by studying the impact of different values of the multiplicative calibration $m$ defined in equation \ref{eq:gamma_sys} and \ref{eq:cmul}.  First, we consider a simple scaling of $m$ with the median redshift, since the redshift can have an impact on galaxy properties, such as size and magnitude, which in turn have an impact on the accuracy with which their shapes are measured.  The redshift scaling is thus taken to be
\begin{equation}
\label{eq:m0}
m_{sys}=m_{0}(1+z_m)^{\beta_m},
\end{equation}

\begin{figure}
\begin{center}
\includegraphics[width=8.5cm,angle=0]{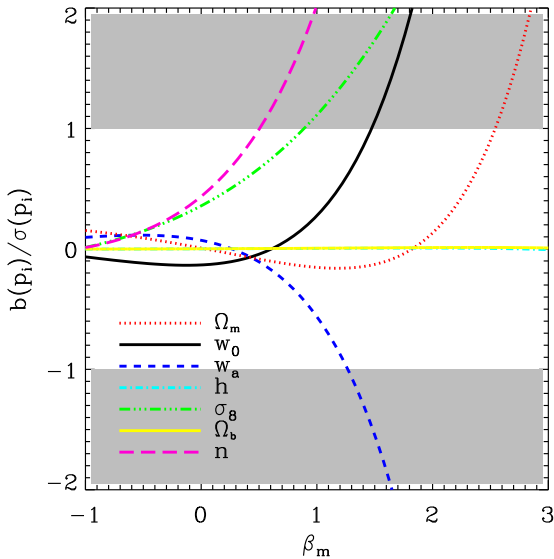}
\caption{The ratio of the bias to statistical error for the multiplicative systematics for the
power-law redshift evolution. The multiplicative calibration parameter is set $m_0 = 5\times10^{-3}$.}
\label{fig:fig7}
\end{center}
\end{figure}

Figure \ref{fig:fig7} shows the fractional bias as a function of $\beta_m$ with $m_0 = 5\times10^{-3}$. If there is no redshift dependence (i.e. if $\beta_m =0$) then setting a tolerance of $m_{0}<m_{tol} = 5\times10^{-3}$ is more than sufficient to keep the biases at a subdominant levels.  Once again, it is not the equation of state parameters that place the most stringent constraints on systematics. In particular, the primordial shape parameter $n$ is the most sensitive.  If one is only concerned with dark energy parameters, we can set a limit of $\beta_m < 1.5$ for this value of $m_{tol}$. Figure \ref{fig:fig7} also shows some distinct features that are not apparent in figure \ref{fig:fig6}, namely that the biases change sign as $\beta_m$ varies. Since $\beta_m$ controls the relative amplitude of the systematics as a function of redshift, this suggests that not all the power spectra are affected in the same way.  For instance, for $w_0$, when the error on the low redshift correlation functions dominates (i.e. for small $\beta_m$) the bias is negative, whereas when the errors on the high redshift correlation functions dominates (i.e. for large $\beta_m$) the bias is positive. This is important since it shows that there exists a trade-off between high and low redshift biases. Moreover, this indicates that if $m$ transitions through zero at some redshift, the multiplicateive systematics are likely to have a greater effect.  This is analogous to what we have found for the additive term. 

\begin{figure}
\begin{center}
\includegraphics[width=8.5cm,angle=0]{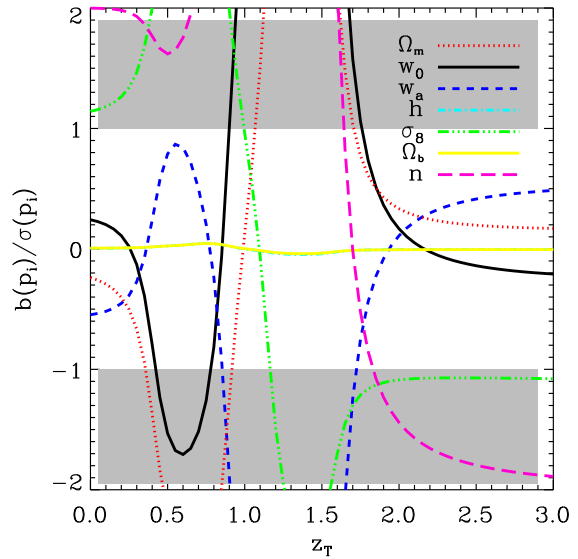}
\caption{The ratio of bias to statistical errors when m is allowed to transition between positive and negative. Here we have set $m_0 = 5\times10^{-3}$ and $\beta_m = 1$.  With the simple redshift evolution used in figure \ref{fig:fig7} one would conclude that this level would be sufficient for the equation of state parameters. Allowing a more complicated redshift evolution can however cause the biases to dominate.}
\label{fig:fig8}
\end{center}
\end{figure}

\begin{figure}
\begin{center}
\includegraphics[width=8.5cm,angle=0]{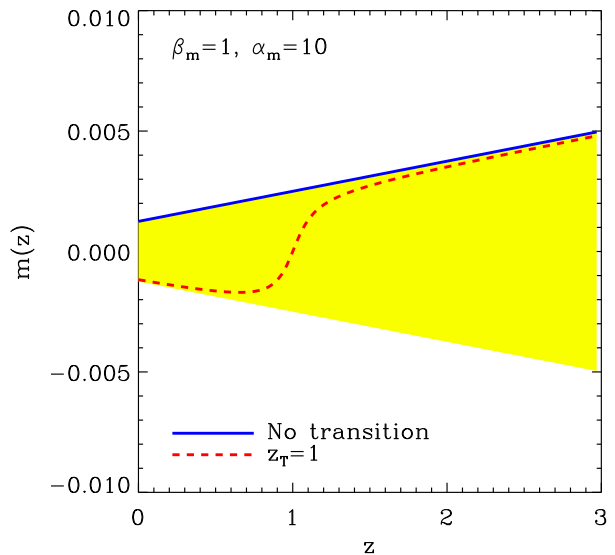}
\caption{The shaded area (yellow) shows the tolerance envelope, with $\beta_m=1$ and $\alpha_=10$. Within this zone the residual systematic can take any form. In blue we show the systematic described in equation \ref{eq:m0} and the red dashed curve shows the systematic from equation \ref{eq:m0tran}.}
\label{fig:fig9}
\end{center}
\end{figure}

\begin{figure}
\begin{center}
\includegraphics[width=8.5cm,angle=0]{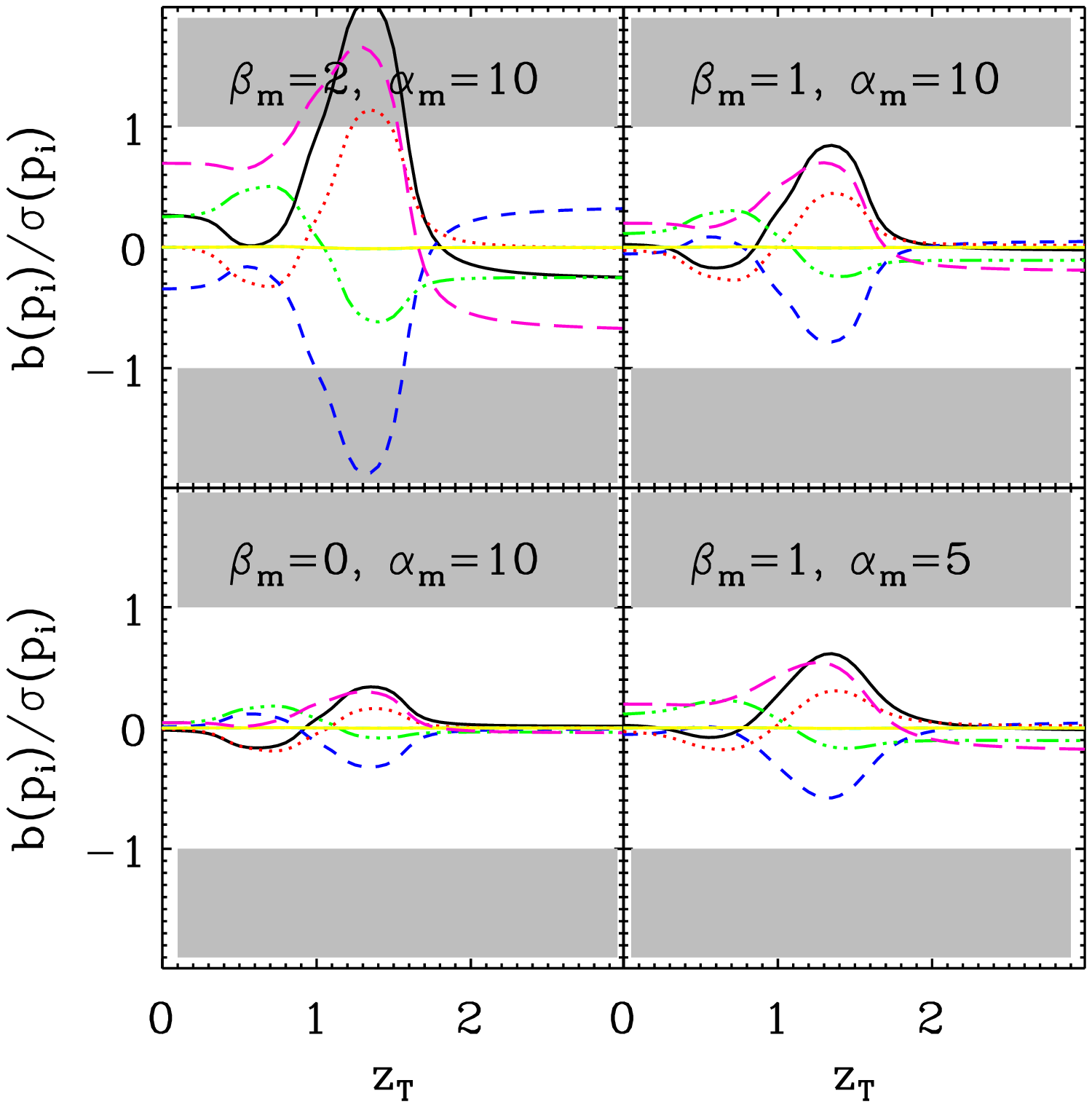}
\caption{Same as Results in figure \ref{fig:fig9} but for different values of $\beta_m$ and $\alpha_m$.  We have set $m_0 = 10^{-3}$, i.e. 5 times more stringent than those iinfered from a simple scaling that is always positive.}
\label{fig:fig10}
\end{center}
\end{figure}

From figure \ref{fig:fig7} we would thus expect that if we are able to constrain $m_0$ to $5\times10^{-3}$ with $\beta_m = 1$, we would expect that the bias on the dark energy parameters would be significantly subdominant to the statical errors. Figure \ref{fig:fig8}, however, tells a different story. 
For any experiment the target is for $m$ to be as close to zero as possible.  Since $m$ can be either positive or negative (or infact transition from one to the other) one can image defining a tolerance envelope such that the unknown systematic can be anywhere in this region.  This is illustrated in figure \ref{fig:fig9}.  The shaded area shows the uncertainty envelope enclosed by $m= \pm m_0 (a+z)^{\beta_m}$, with $m_0= 5\times 10^{-3} $ and $\beta_m = 1$. In blue we show the simple model used from equation \ref{eq:m0}, where $m$ is always positive and the red dashed curve, which was used to produce figure \ref{fig:fig8} is for a systematic that has the form,
\begin{equation}
m_{sys}=\frac{2}{\pi}atan\Big(\alpha_m(z_m-z_T)\Big)~m_{0}(1+z_m)^{\beta_m},
\label{eq:m0tran}
\end{equation}
where $z_T$ is the redshift at which the systematic transitions through zero and $\alpha_m$ controls the rate of the transition. Figure \ref{fig:fig10} shows results for a number of $\alpha_m$ and $\beta_m$ values as a function of the transition redshift.   From this we are able to set the requirements that when we allow $m$ to have a zero transition, the tolerance is $\mathbf{m_{0-tol} = 1\times 10^{-3}}$ for $\mathbf{\beta_m < 1.5}$.

\section{Impact on weak lensing surveys}
In \cite{2006astro.ph.10127A} we studied how the geometries and configurations of future lensing surveys can be optimised to derive the strongest cosmological constraints.  
The present results can be used to extend this work, which solely included
statistical errors, to also include the impact of systematic biases. For this purpose, it is
useful to consider the total mean square error of the cosmological parameters $p_i$, which we have defined in equation \ref{eq:MSE}.

\begin{figure}
\begin{center}
\includegraphics[width=8.5cm,angle=0]{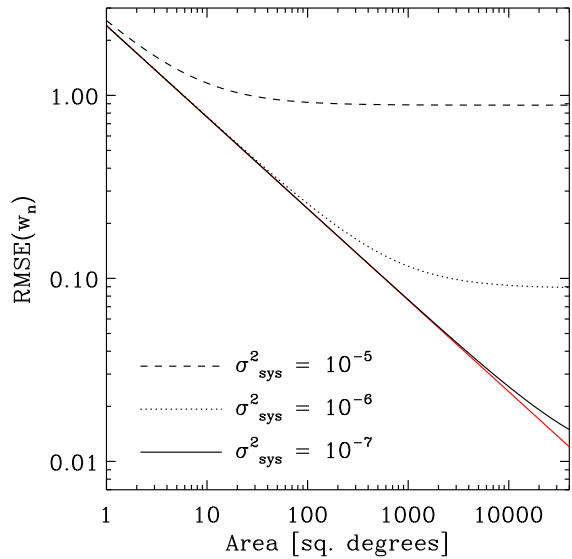}
\caption{The root mean square error (RMSE) on the dark energy equation of state parameter $w_0$ as a function of survey area. This error includes both the effects of statistical errors and the bias induced by a systematics with a shape, $C^{sys}_\ell \propto dC^{33}_\ell/dw_0$ with $\sigma^2_{sys} = 10^{-5}, 10^{-6}~\rm and~10^{-7}$ and $\beta_a=1$. The red line shows the uncertainty with no systematic errors.}
\label{fig:fig_area}
\end{center}
\end{figure}

Figure \ref{fig:fig_area} shows the root mean square error as a function of survey area, with all other properties of the survey set to the
values listed in section \ref{shape_syst}.  Also shown are the impact of different
levels of additive systematics with amplitudes of $\sigma^2_{sys} = 10^{-5}, 10^{-6} ~ \rm and ~10^{-7}$, from top to bottom respectively. The systematics were chosen to have a shape $C^{sys}_\ell \propto dC^{33}_\ell/dw_n$ without redshift evolution.  The figure demonstrates that, when designing future surveys, one needs to consider both the statistical power of the survey and the systematic floor of the survey.  For instance, we see that if an instrument was to have systematics with  $\sigma^2_{sys} = 10^{-5}$ then there is very little point in surveying an area of sky bigger than 20 square degrees.  Alternatively, we see that if we choose to do a 20,000 square degree survey, we will need to control the residual systematic to $\sigma^2_{sys} \sim 10^{-7}$ in order to reach the statistical potential of the survey. 

Since the requirements depend on the ratio of the bias to  statistical errors, the requirements for a specific survey configurations depends on the statistical power of that specific survey.  In \cite{2006astro.ph.10127A} we investigated the statistical power for measuring dark energy of several survey configurations.  We then gave a simple scaling relation for calculation the dark energy Figure of Merit (FoM) as a function of the survey properties.  We have found that these scaling relations hold well and capture most of the relevant physics. In the same spirit we now give the scaling relation for the systematic requirements.  For a survey that has an area $A_s$ , a galaxy surface density $n_g$ (useful for lensing) and a median redshift $z_m$, the tolerance on the additive systematic is,
\begin{equation}
\label{eq:scale1}
\sigma^2_{sys} < 10^{-7} \Bigg(\frac{A_s}{2\times10^4~\rm{deg^2}} \Bigg)^{-0.5} \Bigg(\frac{n_g}{35~\rm{amin^{-2}}}\Bigg)^{-0.5}\Bigg(\frac{z_m}{0.9} \Bigg)^{-0.6} 
\end{equation}
and the requirement on the multiplicative systematic is,
\begin{equation}
\label{eq:scale2}
m_0 <10^{-3} \Bigg(\frac{A_s}{2\times10^4~\rm{deg^2}} \Bigg)^{-0.5} \Bigg(\frac{n_g}{35~ \rm{amin^{-2}}} \Bigg)^{-0.5}\Bigg(\frac{z_m}{0.9} \Bigg)^{-0.6}.
\end{equation}
In both cases we have found that the evolution with redshift must remain weak ($\beta <1.5$).

To illustrate this scaling relation we consider its impact for current day surveys. \cite{2007MNRAS.381..702B} have performed a joint analysis of the latest cosmic shear surveys together
covering about 100 deg$^2$ with a median redshift of ~0.78 and roughly 10 galaxies per arcmin$^2$ on average.  From our scaling relation we see that, for this analysis, the shear calibration $m$ needs to be determined to a precision of better than 3\%. As we have already stated, this is close to the limit of current shear measurement methods in simulated, therefore somewhat idealised,  conditions. This underscores the fact that, for future larger surveys, control of systematic signals needs to be a priority since we are already at the point where they are likely to dominate. With the progress currently being made in the STEP program it is reasonable to expect the shape measurements methods to improve their accuracy for multiplicative systematics, and hence be able to keep up with ever more ambitious surveys. 
For the additive requirement we find that for the \cite{2007MNRAS.381..702B} study we need $\sigma^2_{sys} < 3\times10^{-6}$ and corresponds roughly to an additive error of $c \sim  0.002$ (as defined in STEP \citep{2006MNRAS.368.1323H,2007MNRAS.376...13M}). While less attention has
been given to additive systematics, STEP studies have shown this level is also achievable with
the best current measurement methods, assuming that the measurements are not
limited by PSF modeling and interpolation.
A detailed study of requirements for additive systematics and PSF calibration will be presented in a later paper  (Paulin-Henriksson et al in prep.).

\section{Conclusions}
\label{conclusion}

In this paper, we have outlined a method for computing the biases that residual systematics introduce. This approach involves a simple extension of the Fisher matrix formalism that is now widely used in cosmology to make error forecasts.  As an application, we have used it to study the impact that residual systematic signals will have on future tomographic cosmic shear measurements.  Specifically, we have explored  three different types of shape systematic signal affecting tomographic shear power spectra: (i) additive systematics with no redshift evolution; (ii) additive systematics with redshift evolution; and (iii) multiplicative systematics.  
The requirement target is then to have all types of systematics close to zero.  This defines a tolerance envelope for the systematics that allows the residual systematics errors in the power spectra to be positive or negative within its limits. It is important to note that it is the $worst$ systematic possible within this limit which drives the requirement, not a marginalised average over all systematics. To this end we have investigated a wide class of possible systematic shapes and used the most constraining ones to set our systematic requirements.  In doing this, we have found that, for both the additive and multiplicative parts, it is vital to consider systematics that have positive and negative power spectra $C_\ell^{sys}$.  For instance we see, in the multiplicative case, that investigating only power-law behaviour for its redshift evolution (i.e. $m$ is always positive) can lead to a factor of 5 underestimation of the impact of a systematic within a given tolerance window.  From our calculation we are able to set the following requirements on the survey we have considered (a DUNE-like survey covering 20,000 sq. degrees with 35 galaxies per arcmin$^{-2}$ and a median redshift of $z_m=0.9$):  
\begin{itemize}
\item For both the additive and the multiplicative signals, the redshift evolution needs to be weak $\beta < 1.5$, where the errors  for a given galaxy scale as $(1+z)^\beta$. This is not a trivial requirement since the shapes of more distant galaxies are harder to measure since they are smaller and fainter.
\item The power spectrum of the residual additive shear error, that is the part that is not correlated to the lensing signal, must be controlled such that its amplitude is $\sigma^2_{sys} < 10^{-7}$ (as defined in equation \ref{eq:var}) 
\item The multiplicative part needs to be controlled to a precision of $m_0 <10^{-3}$, where $m_0$ is the shear calibration error.  This means that we need to be able calibrate shears to an accuracy of 0.1\%, which is about one order of magnitude better than the current best measurement methods are able to achieve, as determined by the latest STEP simulations \citep{2006MNRAS.368.1323H,2007MNRAS.376...13M}.
\end{itemize}

These specific requirements apply to our fiducial survey, but we provide scaling relations (Eqs \ref{eq:scale1} and \ref{eq:scale2}) which show the requirements for any survey geometry.  We have shown that for current survey covering $\sim 100$ deg$^{2}$ \citep{2007MNRAS.381..702B}, we need $\sigma^2_{sys} < 3\times10^{-6}$ and $m< 0.03$. This level of accuracy is at the limit of the performance of
the current best shear measurement methods, as demonstrated by STEP. However, further systematics,
such as that arising from PSF calibration and interpolation, are not accounted for by STEP and can
dominate the error budget for future surveys. A discussion of the requirements for additive systematics
and PSF modeling in the context of present and future surveys will be presented in a later paper (Paulin-Henriksson et al., 2007, in prep). 

\newpage
\appendix
\section{shear systematic error and systematic power spectrum}
\label{sec:appA}
The shear at a position can be written as a complex number,
\begin{equation}
\gamma(\theta,z) = \gamma_1(\theta,z) + i \gamma_2(\theta,z).
\end{equation}
This shear field can be measured and used to construct a set of correlation functions between the shear measured at ($\theta,z$) and  ($\theta^\prime,z^\prime$), where $\theta$ and $\theta^\prime$ are the angular positions of two galaxies on the sky with $\theta^\prime=(\theta+\phi)$, and $z$, $z^\prime$ are their respective redshift. The two correlation functions $\xi_+$ and $\xi_-$ can then be constructed as,
\begin{equation}
\xi_+(\phi,z,z^\prime) = \langle \gamma_1(\theta,z) \gamma_1(\theta^\prime,z^\prime)\rangle +  \langle \gamma_2(\theta,z) \gamma_2(\theta^\prime,z^\prime)\rangle,
\end{equation}
and
\begin{equation}
\xi_-(\phi,z,z^\prime) = \langle \gamma_1(\theta,z) \gamma_1(\theta^\prime,z^\prime)\rangle -  \langle \gamma_2(\theta,z) \gamma_2(\theta^\prime,z^\prime)\rangle.
\end{equation}
These real space correlation functions can be combined to calculate the power spectrum between galaxies at a redshifts of $z$ and $z^\prime$,
\begin{equation}
C(\ell,z,z^\prime) = \int \phi d\phi\big(\xi_+(\phi,z,z^\prime)J_0(\ell\phi)  - \xi_-(\phi,z,z^\prime)J_4(\ell\phi) \big),
\label{eq:cl}
\end{equation}
where $J_0$ and $J_4$ are the zeroth and the fourth order Bessel functions. In weak lensing tomography, where galaxies are divided into redshift slices, we use the notation $C_\ell^{ij}$, the correlation function between slices at redshift $z^i$ and $z^j$, for $C(\ell,z,z^\prime)$.

We assume that the observed shear (without noise) is a combination of shear due to lensing plus a residual systematic, $\gamma^{obs} = \gamma^{lens}+\gamma^{sys}$, and that the lensing systematics can be decomposed into an additive error and a multiplicative error,
\begin{equation}
\gamma^{sys}(\theta,z) = \gamma^{add}(\theta,z) + m(z)\gamma^{lens}(\theta,z).
\label{eq:gamma_sys}
\end{equation}
This is similar to the decomposition used by the STEP collaboration \citep{2006MNRAS.368.1323H,2007MNRAS.376...13M}.  Putting this systematic into equation \ref{eq:cl} we see that the systematic power spectrum is composed of two parts, an additive part ($C^{add}$) and a multiplicative powerspectrum ($C^{mul}$),
\begin{equation}
C^{sys}(\ell,z,z^\prime)=C^{add}(\ell,z,z^\prime) + C^{mul}(\ell,z,z^\prime),
\label{eq:cadd_cmul}
\end{equation}
where the additive power spectrum can be calculated directly from the correlation function of $\gamma^{add}$,
\begin{equation}
C^{add}(\ell,z,z^\prime) = \int \phi d\phi\big(\xi^{add}_+(\phi,z,z^\prime)J_0(\ell\phi)  - \xi^{add}_-(\phi,z,z^\prime)J_4(\ell\phi) \big)
\end{equation}
and the multiplicative part is dependent on the lensing power spectrum,
\begin{equation}
C^{mul}(\ell,z,z^\prime) =\big( m(z) + m(z^\prime)\big)C^{lens}(\ell,z,z^\prime).
\label{eq:cmul}
\end{equation}

\newpage
\section{Tabulated bias results}
\label{apx:b}
In this section we give the results for the bias calculation of several examples of systematics that have the same shape as the lensing signal and systematics that resemble $dC_\ell/dp$.  As with the results shown in figure \ref{fig:fig3}, we calculate the bias in each parameter for systematic signals that have an an amplitude given by $\sigma_{\rm sys}^2 = 10^{-7}$.  Table \ref{tbl:1} shows the results for systematic in equation \ref{eq:cl} (i.e. the same shape as the power spectrum shown in the top right panel of figure \ref{fig:fig2}), table \ref{tbl:2} are for systematics in equation \ref{eq:dcldp} where we specifically look at changes in the power spectrum corresponding to small changes in $w_0$.  These systematics are shown in the bottom right panel of figure \ref{fig:fig2}. Finally, table \ref{tbl:3} shows other examples of systematics that are described by  \ref{eq:dcldp}. In this case these are the systematics shown in the bottom left panel on figure \ref{fig:fig2}
\begin{table*}
  \centering 
\begin{tabular}[p]{c|c|c|c|c|c|c|c}
\hline
 & $\frac{b(\Omega_m)}{\sigma(\Omega_m)}$& $\frac{b(w_0)}{\sigma(w_0)}$& $\frac{b(w_a)}{\sigma(w_a)}$& $\frac{b(h)}{\sigma(h)}$ & $\frac{b(\sigma_8)}{\sigma(\sigma_8)}$& $\frac{b(\Omega_b)}{\sigma(\Omega_b)}$& $\frac{b(n)}{\sigma(n)}$ \\
 \hline
$C^{sys}_\ell \propto C^{11}_\ell$ & 0.762 & 0.689 & -0.538 & -0.006 & -0.716 & -0.004 & 0.077\\
$C^{sys}_\ell \propto C^{12}_\ell$ & 0.603 & 0.546 & -0.414 & -0.012 & -0.563 & -0.010 & 0.065\\
$C^{sys}_\ell \propto C^{13}_\ell$ & 0.534 & 0.482 & -0.364 & -0.012 & -0.499 & -0.010 & 0.073\\
$C^{sys}_\ell \propto C^{14}_\ell$ & 0.488 & 0.439 & -0.333 & -0.011 & -0.457 & -0.009 & 0.085\\
$C^{sys}_\ell \propto C^{15}_\ell$ & 0.445 & 0.399 & -0.306 & -0.009 & -0.419 & -0.007 & 0.099\\
$C^{sys}_\ell \propto C^{22}_\ell$ & 0.696 & 0.630 & -0.486 & -0.009 & -0.652 & -0.007 & 0.068\\
$C^{sys}_\ell \propto C^{23}_\ell$ & 0.680 & 0.616 & -0.473 & -0.010 & -0.636 & -0.008 & 0.067\\
$C^{sys}_\ell \propto C^{24}_\ell$ & 0.671 & 0.608 & -0.466 & -0.010 & -0.628 & -0.008 & 0.066\\
$C^{sys}_\ell \propto C^{25}_\ell$ & 0.664 & 0.602 & -0.461 & -0.011 & -0.621 & -0.009 & 0.065\\
$C^{sys}_\ell \propto C^{33}_\ell$ & 0.574 & 0.520 & -0.393 & -0.013 & -0.536 & -0.010 & 0.067\\
$C^{sys}_\ell \propto C^{34}_\ell$ & 0.562 & 0.508 & -0.384 & -0.013 & -0.524 & -0.011 & 0.068\\
$C^{sys}_\ell \propto C^{35}_\ell$ & 0.552 & 0.499 & -0.376 & -0.013 & -0.515 & -0.011 & 0.069\\
$C^{sys}_\ell \propto C^{44}_\ell$ & 0.514 & 0.464 & -0.350 & -0.012 & -0.481 & -0.010 & 0.077\\
$C^{sys}_\ell \propto C^{45}_\ell$ & 0.501 & 0.452 & -0.341 & -0.012 & -0.468 & -0.010 & 0.079\\
$C^{sys}_\ell \propto C^{55}_\ell$ & 0.469 & 0.422 & -0.321 & -0.010 & -0.440 & -0.009 & 0.090\\

\hline
\end{tabular}
  \caption{The ratio of bias to marginalised statistical errors for each parameter from a systematic that has the same shape as the lensing correlation function and $\mathbf{\sigma_{\rm sys}^2 = 4\times10^{-7}}$.  These are the systematic signals shown in the top right panel of figure \ref{fig:fig2}.}
  \label{tbl:1}
\label{tbl:1}
\end{table*}

\begin{table*}
  \centering 
\begin{tabular}[h]{c|c|c|c|c|c|c|c}
\hline
 & $\frac{b(\Omega_m)}{\sigma(\Omega_m)}$& $\frac{b(w_0)}{\sigma(w_0)}$& $\frac{b(w_a)}{\sigma(w_a)}$& $\frac{b(h)}{\sigma(h)}$ & $\frac{b(\sigma_8)}{\sigma(\sigma_8)}$& $\frac{b(\Omega_b)}{\sigma(\Omega_b)}$& $\frac{b(n)}{\sigma(n)}$ \\
 \hline

$C^{sys}_\ell \propto dC^{33}_\ell/\Omega_m$    & 0.557 & 0.466 & -0.330 & -0.015 & -0.416 & -0.014 & 0.070\\
$C^{sys}_\ell \propto dC^{33}_\ell/dw_0$        & -1.325 & -1.089 & 0.885 & -0.026 & 1.034 & -0.028 & -0.134\\
$C^{sys}_\ell \propto dC^{33}_\ell/dw_a$        & -0.188 & -0.150 & 0.171 & -0.030 & 0.163 & -0.030 & 0.072\\
$C^{sys}_\ell \propto dC^{33}_\ell/dh$  & 0.237 & 0.203 & -0.111 & -0.026 & -0.166 & -0.026 & 0.050\\
$C^{sys}_\ell \propto dC^{33}_\ell/d\sigma_8$   & 0.639 & 0.537 & -0.381 & -0.015 & -0.475 & -0.015 & 0.042\\
$C^{sys}_\ell \propto dC^{33}_\ell/d\Omega_b$   & -0.237 & -0.203 & 0.111 & 0.026 & 0.166 & 0.026 & -0.050\\
$C^{sys}_\ell \propto dC^{23}_\ell/dn$          & 0.186 & 0.160 & -0.090 & -0.022 & -0.133 & -0.021 & 0.126\\
\hline
\end{tabular}
  \caption{The ratio of bias to Fisher matrix errors for each parameter from a systematic that is similar to the change in $C^{33}_\ell$ caused by a small change in each of the cosmological parameters. The systematic correlations all have $\mathbf{\sigma_{\rm sys}^2 = 4\times10^{-7}}$. These are the systematic signals shown in the bottom left panel of figure \ref{fig:fig2}.}
  \label{tbl:2}
\end{table*}

\begin{table*}
  \centering 
\begin{tabular}[h]{c|c|c|c|c|c|c|c}
\hline
 & $\frac{b(\Omega_m)}{\sigma(\Omega_m)}$& $\frac{b(w_0)}{\sigma(w_0)}$& $\frac{b(w_a)}{\sigma(w_a)}$& $\frac{b(h)}{\sigma(h)}$ & $\frac{b(\sigma_8)}{\sigma(\sigma_8)}$& $\frac{b(\Omega_b)}{\sigma(\Omega_b)}$& $\frac{b(n)}{\sigma(n)}$ \\
  \hline

$C^{sys}_\ell \propto dC^{11}_\ell/dw_0$  & -0.492 & -0.386 & 0.409 & -0.064 & 0.422 & -0.065 & -0.193\\
$C^{sys}_\ell \propto dC^{12}_\ell/dw_0$  & -0.763 & -0.615 & 0.577 & -0.059 & 0.625 & -0.060 & -0.179\\
$C^{sys}_\ell \propto dC^{13}_\ell/dw_0$  & -0.794 & -0.642 & 0.595 & -0.057 & 0.647 & -0.058 & -0.168\\
$C^{sys}_\ell \propto dC^{14}_\ell/dw_0$  & -0.789 & -0.638 & 0.590 & -0.055 & 0.643 & -0.056 & -0.160\\
$C^{sys}_\ell \propto dC^{15}_\ell/dw_0$  & -0.756 & -0.611 & 0.568 & -0.054 & 0.617 & -0.055 & -0.150\\
$C^{sys}_\ell \propto dC^{22}_\ell/dw_0$  & -1.202 & -0.985 & 0.833 & -0.042 & 0.949 & -0.043 & -0.149\\
$C^{sys}_\ell \propto dC^{23}_\ell/dw_0$  & -1.274 & -1.046 & 0.869 & -0.035 & 1.000 & -0.036 & -0.138\\
$C^{sys}_\ell \propto dC^{24}_\ell/dw_0$  & -1.276 & -1.048 & 0.866 & -0.033 & 1.000 & -0.034 & -0.129\\
$C^{sys}_\ell \propto dC^{25}_\ell/dw_0$  & -1.250 & -1.027 & 0.849 & -0.032 & 0.980 & -0.033 & -0.118\\
$C^{sys}_\ell \propto dC^{33}_\ell/dw_0$  & -1.325 & -1.089 & 0.885 & -0.026 & 1.034 & -0.028 & -0.134\\
$C^{sys}_\ell \propto dC^{34}_\ell/dw_0$  & -1.287 & -1.058 & 0.854 & -0.023 & 1.003 & -0.025 & -0.130\\
$C^{sys}_\ell \propto dC^{35}_\ell/dw_0$  & -1.286 & -1.056 & 0.852 & -0.023 & 1.002 & -0.025 & -0.125\\
$C^{sys}_\ell \propto dC^{44}_\ell/dw_0$  & -1.185 & -0.973 & 0.782 & -0.020 & 0.923 & -0.022 & -0.130\\
$C^{sys}_\ell \propto dC^{45}_\ell/dw_0$  & -1.177 & -0.966 & 0.777 & -0.022 & 0.918 & -0.023 & -0.134\\
$C^{sys}_\ell \propto dC^{55}_\ell/dw_0$  & -1.187 & -0.973 & 0.787 & -0.025 & 0.928 & -0.026 & -0.145\\
\hline
\end{tabular}
  \caption{The ratio of bias to Fisher matrix errors for each parameter from a systematic that is similar to the change in $C^{ij}_\ell$ corresponding to a small change in $w_0$. As before, we use $\mathbf{\sigma_{\rm sys}^2 = 4\times10^{-7}}$. These are the systematic signals shown in the bottom right panel of figure \ref{fig:fig2}.}
  \label{tbl:3}
\end{table*}

\section*{Acknowledgements}
We thank Stephane Paulin, Sarah Bridle, Lisa Voigt
and the DUNE weak lensing working group members for useful discussions. AA would also like to thank Prof. K. S. Cheng and Dr. T. Harko of Hong Kong University for their kind hospitality. 

\bibliographystyle{astron}
\bibliography{../../mybib}

\end{document}